\documentclass[prb,twocolumn,superscriptaddress,showpacs,amsmath,amssymb,longbibliography]{revtex4-2}
\usepackage{amsmath, amsfonts, amssymb, bm}
\usepackage{float}
\usepackage{mathtools}
\usepackage[inline]{enumitem}
\usepackage{graphicx}
\usepackage{xcolor}
\usepackage[breaklinks,hypertexnames=false]{hyperref}
\hypersetup{colorlinks,linkcolor={magenta},citecolor={blue},urlcolor={blue}}
\usepackage[capitalize]{cleveref}
\usepackage{lineno}

\begin{document}
\title{Orientation-dependent transport in junctions formed by $d$-wave altermagnets and $d$-wave superconductors}
	
\newcommand{\tianjin}{Center for Joint Quantum Studies, Tianjin Key Laboratory of Low Dimensional Materials Physics and Preparing Technology, Department of Physics, Tianjin University, Tianjin 300354, China}

\newcommand{\uam}{Department of Theoretical Condensed Matter Physics, Universidad Aut\'onoma de Madrid, 28049 Madrid, Spain}
\newcommand{\ifimac}{Condensed Matter Physics Center (IFIMAC), Universidad Aut\'onoma de Madrid, 28049 Madrid, Spain}
\newcommand{\inc}{Instituto Nicol\'as Cabrera, Universidad Aut\'onoma de Madrid, 28049 Madrid, Spain}
    
\newcommand{\nagoya}{Department of Applied Physics, Nagoya University, Nagoya 464-8603, Japan}
	
\newcommand{\Okayama}{Faculty of Environmental Life, Natural Science and Technology, Okayama University, 700-8530 Okayama, Japan}
	
\newcommand{\Uppsala}{Department of Physics and Astronomy, Uppsala University, Box 516, S-751 20 Uppsala, Sweden}
	
\author{Wenjun Zhao}
\affiliation{\tianjin}
	
\author{Yuri Fukaya}
\affiliation{\Okayama}
	
\author{Pablo Burset}
\affiliation{\uam}
\affiliation{\ifimac}
\affiliation{\inc}
	
\author{Jorge Cayao}
\affiliation{\Uppsala}

\author{Yukio Tanaka}
\affiliation{\nagoya}
	
\author{Bo Lu}
\affiliation{\tianjin}

\date{\today}
	
\begin{abstract}
We investigate de Gennes-Saint-James states and Josephson effect in hybrid junctions based on $d$-wave altermagnet and $d$-wave superconductor. 
Even though these states are associated to long junctions, we find that the $d_{x^{2}-y^{2}}$-altermagnet in a normal metal/altermagnet/$d$-wave superconductor junction forms de Gennes-Saint-James states in a short junction due to an enhanced mismatch between electron and hole wave vectors. As a result, the zero-bias conductance peak vanishes and pronounced resonance spikes emerge in the subgap conductance spectra. By contrast, the $d_{xy}$-altermagnet only features de Gennes-Saint-James states in the long junction. 
Moreover, the well-known features such as V-shape conductance for $d_{x^2-y^2}$ pairings and zero-biased conductance peak for $d_{xy}$ pairings 
are not affected by the strength of $d_{xy}$-altermagnetism in the short junction. 
We also study the Josephson current-phase relation $I\left(
\varphi \right) $ of $d$-wave superconductor/altermagnet/$d$-wave superconductor hybrids, where $\varphi $ is
the macroscopic phase difference between two $d$-wave superconductors. In symmetric junctions, we obtain anomalous current phase relation such as a $0$-$\pi$ transition by changing either the orientation or the magnitude of the altermagnetic order parameter and dominant higher Josephson harmonics. 
Interestingly, we find the first-order
Josephson coupling in an asymmetric $
d_{x^{2}-y^{2}}$-superconductor/altermagnet/$d_{xy}$-superconductor junction when the symmetry of altermagnetic order parameter is neither $d_{x^{2}-y^{2}}$- nor $
d_{xy}$-wave. We present the symmetry analysis and conclude that the anomalous orientation-dependent current-phase relations are ascribed to the peculiar feature of the altermagnetic spin-splitting field.
\end{abstract}
\maketitle
	
\section{Introduction}

Heterostructures formed by superconductors coupled to normal state materials bear a great interest in condensed matter physics due to their potential for realizing emergent superconducting phenomena of use for future quantum applications \cite{DeFranceschi2010,Linder2015,Frolov2020,Prada2020,ptae065}. These novel states are often characterized by low-energy excitations within the superconducting gap, or subgap states, that can be controlled with great precission. In the simplest case, when a finite size normal metal is in contact with a superconductor, subgap bound states appear in the normal region known as de Gennes-Saint-James (dGSJ) states \cite{DeSaint63}
which are also called Andreev bound states \cite{Mizushima18,Sauls18}.  According to Bohr-Sommerfeld quantization \cite{zagoskin2016quantum}, dGSJ states form when the total accumulation of phase becomes a multiple of $2\pi$ during a complete cycle comprising two Andreev reflections at the normal metal-superconductor interface and two normal reflections at the open edge of the normal metal. The dGSJ states manifest themselves as a series of pronounced conductance spikes \cite{Rowell66} which have been observed experimentally in various metallic materials backed on one side by a superconductor \cite{Rowell73,Tessmer93}. The spikes oscillate as a function of the thickness of the normal region with characteristic length $\xi _{S}=\hbar v_{F}/ (\pi \Delta) $, where $\Delta$ is the superconducting gap. Notably, with unconventional pairing states, such as $d$-wave pairings, the dGSJ states evolve into flat zero-energy surface Andreev bound states (ZESABSs) \cite{Hu94,kashiwaya2000,Lofwander2001,tanaka2012review} and can be found at normal regions of arbitrary thickness \cite{Experiment1,odd3b}. The ZESABSs manifest themselves as a zero bias conductance peak in tunneling spectroscopy 
\cite{TK95,Experiment1,Experiment3,Experiment4,Experiment5,Experiment6,kashiwaya2000}. 

\begin{figure}[t]
\begin{center}
\includegraphics[width=80mm]{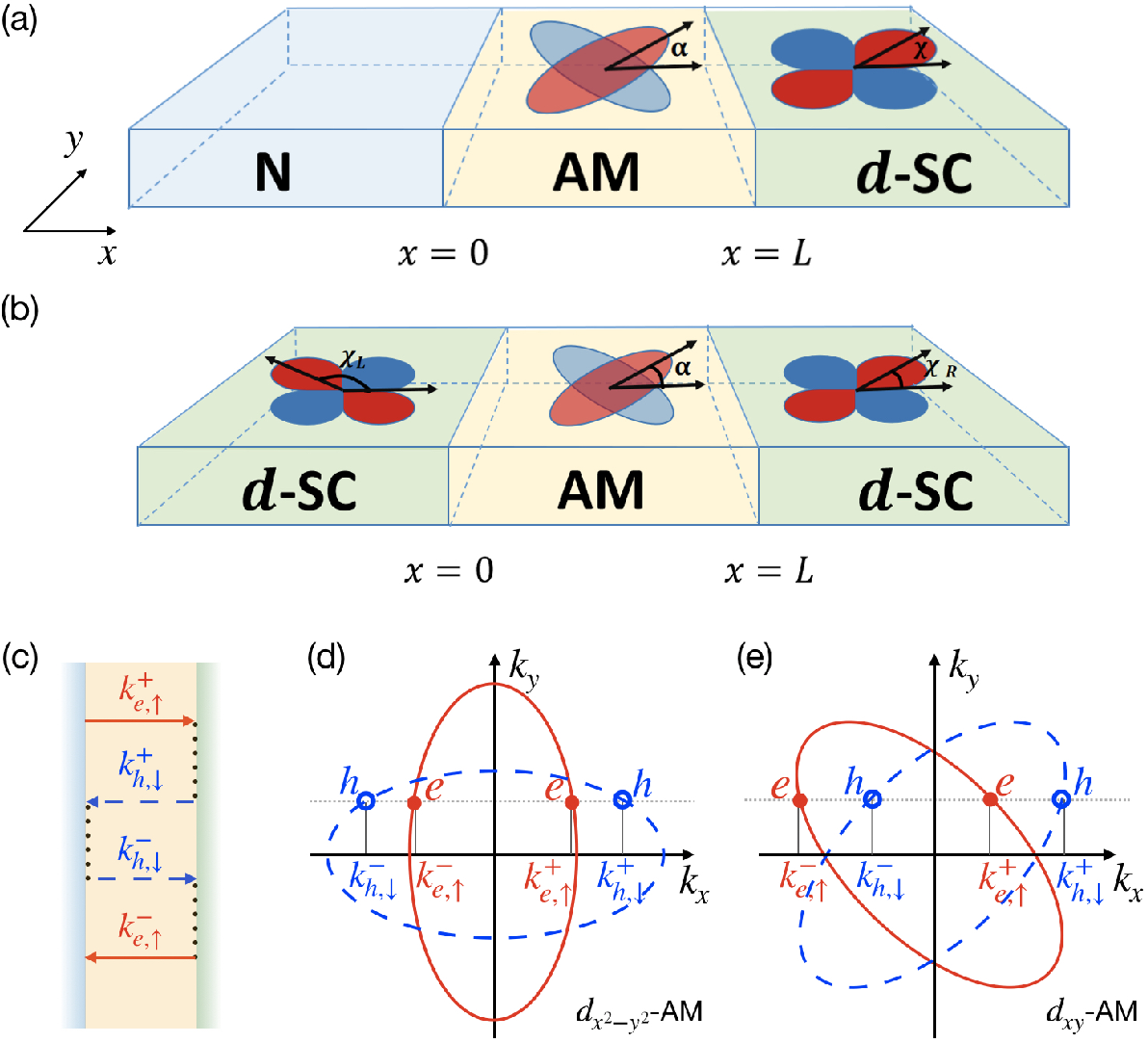}
\end{center}
\caption{Schematics of (a) the normal metal (N)/altermagnet (AM)/$d$-wave superconductor ($d$-SC) junction and (b) the $d$-SC/AM/$d$-SC junction. (c) sequential transport processes inside AM interlayer which are necessary to form dGSJ states. Solid line stands for electron and dashed line stands for hole. (d) For the $d_{x^2-y^2}$-AM case, band splitting enhances phase accumulation $(k_{e,\uparrow}^{+}-k_{e,\uparrow}^{-}+k_{h,\downarrow}^{-}-k_{h,\downarrow}^{+})L$ and thus dGSJ states can be generated with small $L$. Such phase accumulation is almost zero for the $d_{xy}$-AM case (e) and no dGSJ state can be formed in the short junction.}
\label{Fig1}
\end{figure}
	
The dGSJ states have also been extensively studied in ferromagnet/superconductor hybrids
\cite{Zareyan01,Yeyati01,Halterman01,Halterman02,KRAWIEC20037,eschrig18}. In this regard, it was shown that the oscillations at the scale of $\xi _{F}$ in the induced pairing amplitude and in the local density of states at the Fermi energy are related to the evolution of the dGSJ states \cite{Yeyati01}, where  $\xi _{F}=\hbar v_{F}/ (\pi M) $ is the ferromagnetic coherence length with $M$ the exchange field in the ferromagnet. Here, $\xi _{F}$ is generally much smaller than $\xi _{S}$ so that the oscillation is short ranged. For strong ferromagnets such as half metals, Andreev reflection is often suppressed or forbidden 
\cite{Beenakker1995,Kashiwaya1999,Zutic1999}. Remarkably, equal spin Andreev reflection was observed in experiments due to spin-flip and spin-mixing processes at the spin-active boundary between ferromagnet and superconductor \cite{visani2012}, and is the key factor to produce the dGSJ states.
	
Recently, a novel class of magnets dubbed altermagnets (AMs) has attracted substantial interest \cite{NakaNatCommun2019,Hayami19,Ahn2019,NakaPRB2020,LiborSAv,Hayami20,YuanPRB20,MazinPNAS,LiborPRX22,landscape22,BhowalPRX24,Bailing,Mhu2024,jungwirth2024,tamang2024,cheong2024altermagnetism}. AM materials exhibit anisotropic non-relativistic spin splitting field, without net magnetization. From symmetry perspective, the opposite-spin sublattices in AMs are connected by rotational or mirror symmetries, rather than translational or inversion symmetries, leading to even-parity order parameter such as $d$-, $g$-wave, etc. 
Various AM materials have been found like  ${\mathrm{RuO}}_{2}$ \cite{Ahn2019,LiborSAv,LiborPRX22} and ${\mathrm{MnTe}}$ \cite{Lee24,Osumi2024,krempasky2024}, see also Refs. \onlinecite{landscape22,Bailing,tamang2024}.

The interplay between altermagnetism and superconductivity bears fundamental interest and is expected to be useful for spintronic applications. It was found that the Andreev reflection in AM/superconductor (SC) junctions is strongly orientation-dependent \cite{Sun23,Papaj23,maeda24,mondal2024,Niu_2024} and spin-polarized~\cite{nagae2024}. Also, it was shown that even without net magnetization, there are $0$-$\pi$ oscillations when modulating the junction length and AM strength in Josephson junctions with spin-singlet SCs and $d$-wave AMs \cite{Ouassou23,Beenakker23,zhang2024}. The exotic $\varphi$-Josephson junction was also predicted in altermagnetic Josephson junctions with simplest $s$-wave pairing potential~\cite{Bo2024, fukaya2024}. It was shown that the current-phase relation has a rich diversity of anomalous characteristics such as multiple nodes~\cite{Bo2024}, tunable skewness~\cite{sun2024}, and orientation dependence \cite{Cheng24}. Despite the recent efforts, transport in junctions formed by AMs and high-temperature superconductors has so far received little attention. Given that $d$-wave AMs represent the magnetic counterparts of high-temperature superconductor with $d$-wave pairings, it is natural to wonder about transport in junctions formed by them. This problem, however, has not been addressed yet.
 
In this paper, we focus on the transport characteristics in two types of hybrid junctions combining $d$-wave superconductors ($d$-SCs) and a $d$-AM interlayer.	
First, we study the normal metal (N)/AM/$d$-SC junction as depicted in \cref{Fig1}(a). We investigate the possibility of the formation of dGSJ states and the robustness of the ZESABS against altermagnetic order. We demonstrate the schematics of the quasiparticle trajectories in \cref{Fig1}(c) with four modes to form dGSJ states, which are depicted in Figs. \ref{Fig1}(d)(e). The phase accumulation along the $x$-direction for a fixed transverse momentum is proportional to $(k_{e,\uparrow}^{+}-k_{e,\uparrow}^{-}+k_{h,\downarrow}^{-}-k_{h,\downarrow}^{+})L$ where  $L$ is the width of the AM interlayer. Due to the distinct band splitting, the phase accumulation is enhanced by $d_{x^2-y^2}$-AM order but largely vanishes for $d_{xy}$-AM order. Thus, the dGSJ states can be formed in the short junction with $d_{x^2-y^2}$-AM order and the oscillatory transport behavior can thus be expected. Interestingly, we found that the phase accumulation also affects the formation of ZESABSs, which are vulnerable to the $d_{x^2-y^2}$-AM order but almost immune to the $d_{xy}$-AM order, though the time-reversal symmetries are broken for both cases. Second, we investigate the Josephson effect of the $d$-SC/AM/$d$-SC Josephson junction as shown in Fig. \ref{Fig1}(b). We calculate the Josephson current $I(\varphi)$ for various orientations of the junctions where $\varphi$ is the macroscopic phase difference between two $d$-SCs. We obtain the altermagnetism-induced $0$-$\pi$ transitions in symmetric junctions. In the case of $d_{x^{2}-y^{2}}$-SC/AM/$d_{xy}$-SC junction, where the first order of Josephson current is absent without AM \cite{Yip1995,tanaka971}, we find that the first-order Josephson coupling reemerges when AM is neither $d_{x^{2}-y^{2}}$- nor $d_{xy}$-wave. We further provide the explanation of our numerical results by symmetry analysis.

The paper is organized as follows: In Sec.~\ref{sec2}, we introduce our model and formalism. In Sec.~\ref{sec3}, we show numerical results for N/AM/$d$-SC junctions and discuss the dGSJ states. In Sec.~\ref{sec4}, we show the Josephson effect in $d$-SC/AM/$d$-SC junctions. Our conclusions are given in Sec.~\ref{sec5}.

\section{Model and Formalism}
\label{sec2}

In this section, we provide a formulation to calculate conductance and
Josephson current using the scattering approach. As depicted in Figs.~\ref%
{Fig1}(a) and \ref{Fig1}(b), we consider N/AM/$d$-SC and $d$-SC/AM/$d$-SC
junctions which are translation invariance in $y$ direction. $\hat{H}$
corresponds to the Hamiltonian of low-energy excitations
\begin{equation}
\hat{H}=\left(
\begin{array}{cc}
H_{0} & \hat{\Delta} \\
\hat{\Delta} & -H_{0}^{\ast }%
\end{array}%
\right) ,\hat{\Delta}=i\hat{\sigma}_{y}\Delta ,
\end{equation}%
\begin{equation}
H_{0}=\frac{\hbar ^{2}\mathbf{k}^{2}}{2m}+U-\mu +\mathcal{M}\hat{\sigma}_{z},
\end{equation}%
in the basis $(\psi _{\uparrow },\psi _{\downarrow },\psi _{\uparrow }^{\dag
},\psi _{\downarrow }^{\dag })^{T}$. $\Delta $ is the position-dependent $d$%
-wave pairing potential. The wave vector $\mathbf{k}$ is given by $\mathbf{k}%
=(k_{x},k_{y})$ and $\mu $ is the uniform chemical potential so that the
Fermi wave vector is $k_{F}=\sqrt{2m\mu }/\hbar $, with $m$ the electron mass. $U$ is the barrier
potential at the left boundary $U\left( x\right) =U_{1}\delta \left(
x\right) $ and we define a dimensionless parameter $Z=mU_{1}/(\hbar
^{2}k_{F})$. $\hat{\sigma}_{i=x,y,z}$ are Pauli matrices in the spin space. $%
\mathcal{M}$ denotes the exchange potential of altermagnet and without loss
of generality, the N\'{e}el vector of AM is along $z$-axis,%
\begin{equation}
\mathcal{M}=\left[ \frac{J_{1}}{2}\left( k_{x}^{2}-k_{y}^{2}\right)
+J_{2}k_{x}k_{y}\right] \Theta (x)\Theta (L-x),
\end{equation}%
with $\Theta $ being the Heaviside
function, $J_{1}=2Jk_{F}^{-2}\sin 2\alpha $, $J_{2}=2Jk_{F}^{-2}\cos 2\alpha $
and $J$ the strength of the exchange energy of the AM. The junction length is $L$%
. We denote $\alpha $ the angle between the lobe of the direction of
altermagnet and $x$-axis. For $\alpha =0$, the magnetization has pure $%
d_{x^{2}-y^{2}}$-wave symmetry and for $\alpha =\pi /4$, it has pure $d_{xy}$%
-wave symmetry. To find the conductance and Josephson current, we construct the wave functions in
each region of the junction.


In the N/AM/$d$-SC junction as shown in~\cref{Fig1}(a), $\Delta $ is given by
\begin{equation}
\Delta =\Delta _{0}\cos \left( 2\theta -2\chi \right) \Theta (x-L),
\end{equation}%
where $\theta $ is the propagating angle in superconductors of
quasiparticles with $k_{y}=k_{F}\sin \theta $. The quantity $\chi $ is taken to be the angle of the positive $d$%
-wave lobe with respect to the interface normal. We denote $\psi _{1(2)}$
for wave functions as an incident spin-$\uparrow $($\downarrow $) electron
with energy $E$ injects from the normal side. Due to the translational
invariance along the $y$-axis, the transverse momentum $k_{y}$ is conserved.
On the normal side, we have%
\begin{widetext}
\begin{equation}
\psi _{1\left( 2\right) }\left( x\leq 0\right) =\left( e^{ik^{+}x}\check{e}%
_{1\left( 2\right) }+a_{1\left( 2\right) }e^{ik^{-}x}\check{e}_{4\left(
3\right) }+b_{1\left( 2\right) }e^{-ik^{+}x}\check{e}_{1\left( 2\right)
}\right) e^{ik_{y}y}.
\end{equation}%
Here, we denote $k^{\pm }=\sqrt{2m\left( \mu \pm E\right) /\hbar
^{2}-k_{y}^{2}}$ as the wave vectors for electrons ($+$) and holes ($-$), and $a_{i}$ and $b_{i}
$ are the coefficients of reflected waves. We define $\check{e}%
_{1}=(1,0,0,0)^{T}$, $\check{e}_{2}=(0,1,0,0)^{T}$, $\check{e}%
_{3}=(0,0,1,0)^{T}$ and $\check{e}_{4}=(0,0,0,1)^{T}$ as basis functions. In
the middle AM region, we have%
\begin{equation}
\psi _{1\left( 2\right) }\left( 0<x<L\right) =\left( w_{1\left( \bar{1}%
\right) }e^{ik_{e,\uparrow \left( \downarrow \right) }^{+}x}\check{e}%
_{1\left( 2\right) }+w_{2\left( \bar{2}\right) }e^{ik_{e,\uparrow \left(
\downarrow \right) }^{-}x}\check{e}_{1\left( 2\right) }+w_{3\left( \bar{3}%
\right) }e^{ik_{h,\downarrow \left( \uparrow \right) }^{+}x}\check{e}%
_{4\left( 3\right) }+w_{4\left( \bar{4}\right) }e^{ik_{h,\downarrow \left(
\uparrow \right) }^{-}x}\check{e}_{4\left( 3\right) }\right) e^{ik_{y}y},
\end{equation}%
with wave vectors%
\begin{eqnarray}
k_{e,s}^{\pm } &=&\pm \frac{\hbar }{\hbar ^{2}+p_{s}mJ_{2}}\sqrt{2m\left(
\mu +E\right) \left( 1+\frac{mJ_{2}}{\hbar ^{2}}\right) -\hbar ^{2}k_{y}^{2}+%
\frac{m^{2}\left( J_{1}^{2}+J_{2}^{2}\right) k_{y}^{2}}{\hbar ^{2}}}-\frac{%
p_{s}mJ_{1}k_{y}}{\hbar ^{2}+p_{s}mJ_{2}}, \\
k_{h,s}^{\pm } &=&\pm \frac{\hbar }{\hbar ^{2}+p_{s}mJ_{2}}\sqrt{2m\left(
\mu -E\right) \left( 1+\frac{mJ_{2}}{\hbar ^{2}}\right) -\hbar ^{2}k_{y}^{2}+%
\frac{m^{2}\left( J_{1}^{2}+J_{2}^{2}\right) k_{y}^{2}}{\hbar ^{2}}}-\frac{%
p_{s}mJ_{1}k_{y}}{\hbar ^{2}+p_{s}mJ_{2}}.
\end{eqnarray}%
Here, we define $p_{s=\uparrow }=+1$ and $p_{s=\downarrow }=-1$, while $w_{i}$ and
$w_{\bar{i}}$ are the coefficients of scattering waves. On the superconducting side, the wave functions are given by
\begin{equation}
\psi _{1\left( 2\right) }\left( x\geq L\right) =\left( f_{1\left( \bar{1}%
\right) }e^{iq^{+}x}\check{e}_{1\left( 2\right) }\pm f_{1\left( \bar{1}%
\right) }\gamma _{1}e^{iq^{+}x}\check{e}_{4\left( 3\right) }+g_{1\left( \bar{%
1}\right) }\gamma _{2}e^{-iq^{-}x}\check{e}_{1\left( 2\right) }\pm
g_{1\left( \bar{1}\right) }e^{-iq^{-}x}\check{e}_{4\left( 3\right) }\right)
e^{ik_{y}y},
\end{equation}%
where the wave vectors are $q^{\pm }=\sqrt{2m\left( \mu
\pm \sqrt{E^{2}-\Delta ^{2}}\right) /\hbar ^{2}-k_{y}^{2}}$, and $f_{1}$, $f_{%
\bar{1}}$, $g_{1}$ and $g_{\bar{1}}$ are the coefficients. The coherence
factors $\gamma _{1}$ and $\gamma _{2}$ are as follows
\begin{equation}
\gamma _{1}=\frac{\Delta (T)\cos \left( 2\theta -2\chi \right) }{E+\sqrt{%
E^{2}-\Delta (T)^{2}\cos ^{2}\left( 2\theta -2\chi \right) }},\gamma _{2}=%
\frac{\Delta (T)\cos \left( 2\theta +2\chi \right) }{E+\sqrt{E^{2}-\Delta
(T)^{2}\cos ^{2}\left( 2\theta +2\chi \right) }}.
\end{equation}%
The scattering coefficients are determined by continuity of wave functions $%
\psi \big|_{x=0^{+}}=\psi \big|_{x=0^{-}}$, $\psi \big|_{x=L^{+}}=\psi \big|%
_{x=L^{-}}$ and
\begin{eqnarray}
\left( \frac{\hbar ^{2}}{m}+J_{2}\hat{\sigma}_{z}\right) \partial _{x}\psi %
\big|_{x=0^{+}}-\frac{\hbar ^{2}}{m}\partial _{x}\psi \big|_{x=0^{-}}
&=&\left( -iJ_{1}k_{y}\hat{\sigma}_{z}+2U_{1}\right) \psi \big|_{x=0}, \\
\frac{\hbar ^{2}}{m}\partial _{x}\psi \big|_{x=L^{+}}-\left( \frac{\hbar ^{2}%
}{m}+J_{2}\hat{\sigma}_{z}\right) \partial _{x}\psi \big|_{x=L^{-}}
&=&\left( iJ_{1}k_{y}\hat{\sigma}_{z}\right) \psi \big|_{x=L}.
\end{eqnarray}%
Here, we consider no barrier between AM and $d$-SC but assume a large barrier
between N and AM. Thus the voltage drop across N/AM boundary is much larger
than that across AM/$d$-SC boundary. As a result, we can utilize the
Blonder-Tinkham-Klapwijk (BTK) formalism \cite{BTK82,TK95} to compute the
differential conductance \cite{Olthof18}
\begin{equation}
\sigma /\sigma _{0}=\int_{-\pi /2}^{\pi /2}\sigma \left( \theta \right) \cos
\theta d\theta \Big/\int_{-\pi /2}^{\pi /2}\sigma _{0}\left( \theta \right)
\cos \theta d\theta ,  \label{cond1}
\end{equation}%
with $\sigma \left( \theta \right) =2+\left\vert a_{1}\right\vert
^{2}+\left\vert a_{2}\right\vert ^{2}-\left\vert b_{1}\right\vert
^{2}-\left\vert b_{2}\right\vert ^{2}$. Here, $\sigma _{0}$ denotes the
conductance when the superconductor is in the normal state $\sigma
_{0}\left( \theta \right) =2-\left\vert b_{N1}\right\vert ^{2}-\left\vert
b_{N2}\right\vert ^{2}$, and $b_{N1(2)}$ is the corresponding scattering
coefficient for the spin-$\uparrow $ ($\downarrow $) electron reflection.

\begin{figure*}[tbp]
\includegraphics[width=18cm]{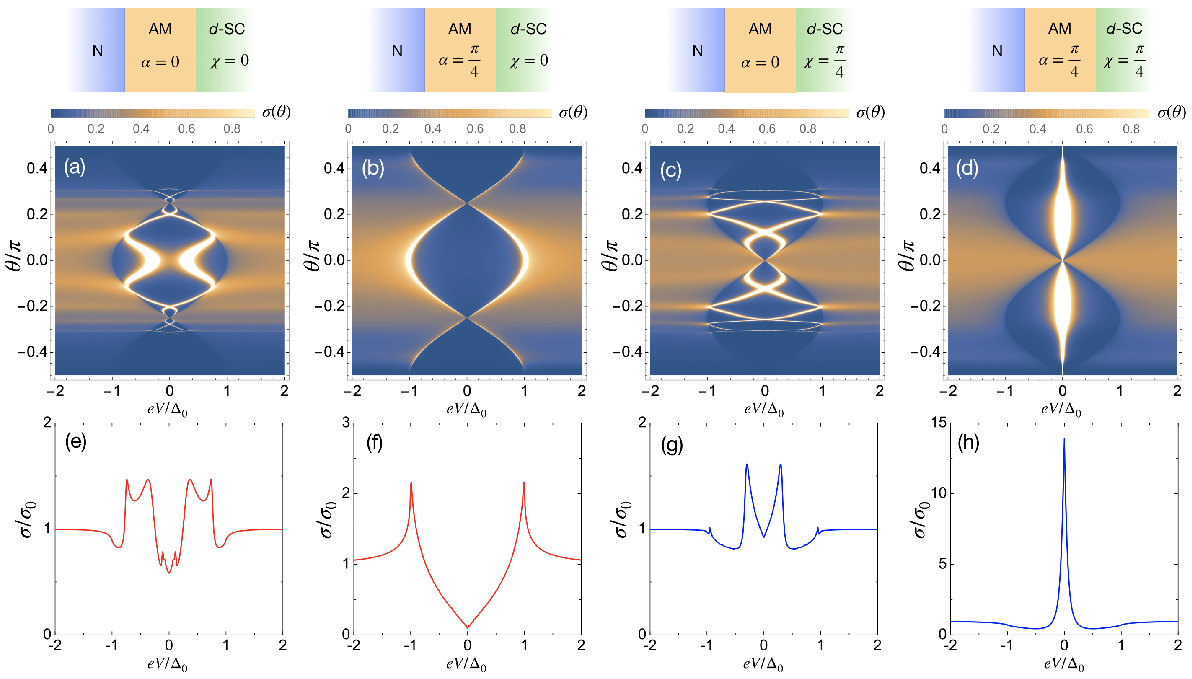}
\caption{(a)-(d) Angle-resolved conductance $\sigma (\theta )
$. We set $Z=2$, $k_{F}L=10$ and $J/\mu =0.2$ for all panels. The
superconductor has $d_{x^{2}-y^{2}}$-wave symmetry ($\chi =0$) in
panels (a)(b) and has $d_{xy}$-wave symmetry ($\chi =\pi /4$%
) in panels (c)(d). The symmetry of altermagnetic order parameter is $%
d_{x^{2}-y^{2}}$-wave ($\alpha =0$) in panels (a)(c) and $d_{xy}$%
-wave ($\alpha =\pi /4$) in panels (b)(d). (e)-(h) are the
total conductance corresponding to (a)-(d). }
\label{Fig2}
\end{figure*}


For the $d$-SC/AM/$d$-SC Josephson junction as shown in~\cref{Fig1}(b), the pair potential is given by
\cite{TK96a,tanaka961,tanaka971,Barash}
\begin{equation}
\Delta \left( x\right) =\left\{
\begin{array}{ll}
\Delta \left( T\right) \cos \left( 2\theta -2\chi _{L}\right) e^{i\varphi },
& x<0, \\
0, & 0<x<L, \\
\Delta \left( T\right) \cos \left( 2\theta -2\chi _{R}\right) , & x>L.%
\end{array}%
\right.
\end{equation}%
Here, $\varphi $ is the macroscopic phase difference between the left and
right superconductors and $\chi _{L}$ and $\chi _{R}$ are the angles of the
positive $d$-wave lobe on the left and right side, respectively. The pair
potential at zero temperature $\Delta (T=0)$ is still $\Delta _{0}$ and its
temperature dependence is determined by mean-field approximation \cite%
{tanaka961,tanaka971}. We focus on the Josephson effect in the low
temperature limit in this paper. Now, we can solve wave functions of each
region for the scattering processes. By using the standard
Furusaki-Tsukada's formula \cite{FT91,tanaka961,tanaka971}, we obtain the Josephson current
\begin{equation}
I=\int_{-\pi /2}^{\pi /2}\frac{ek_{B}T}{2\hbar }\sum\limits_{\omega
_{n},s=\uparrow ,\downarrow }\left( \frac{\Delta \left( T\right) \cos \left(
2\theta -2\chi _{L}\right) a_{he,s}}{\sqrt{\omega _{n}^{2}+\Delta ^{2}\left(
T\right) \cos ^{2}\left( 2\theta -2\chi _{L}\right) }}-\frac{\Delta \left(
T\right) \cos \left( 2\theta +2\chi _{L}\right) a_{eh,s}}{\sqrt{\omega
_{n}^{2}+\Delta ^{2}\left( T\right) \cos ^{2}\left( 2\theta +2\chi
_{L}\right) }}\right) \cos \theta d\theta ,  \label{FT}
\end{equation}%
\begin{figure*}[tbp]
\includegraphics[width=18cm]{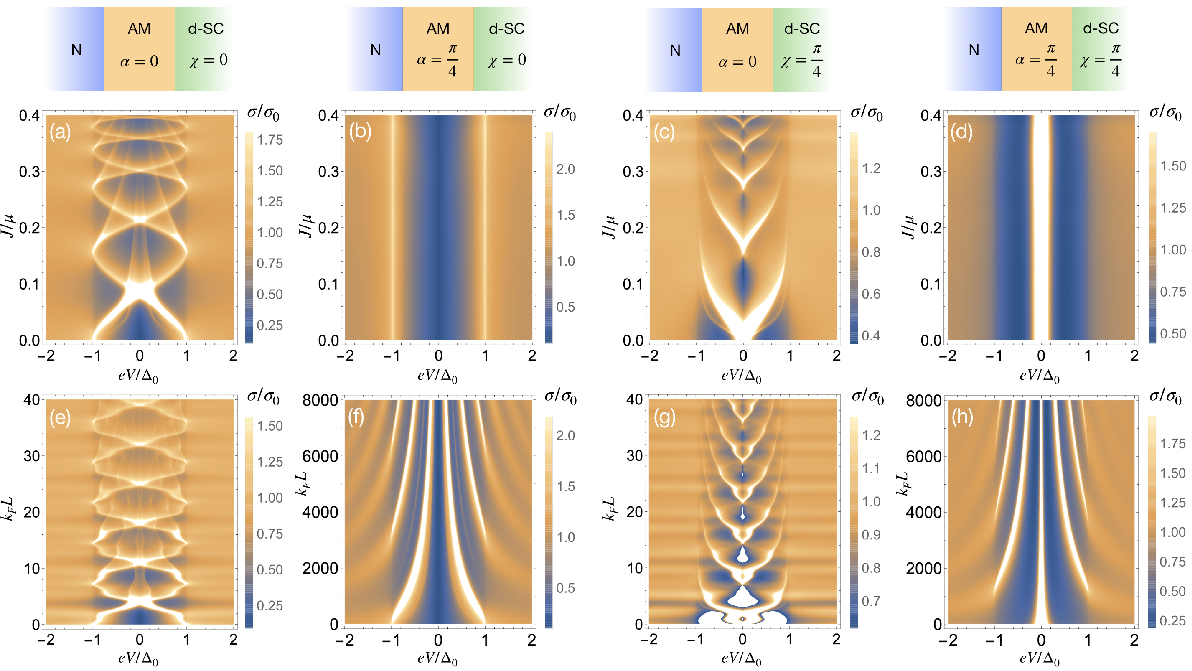}
\caption{Parameter dependence of conductance. (a)-(d) conductance varies
with $J$ and $eV$ for $k_{F}L=10$. (e)-(h) conductance varies with $L$ and $%
eV$ for $J/\mu =0.2$. The AM order is $d_{x^{2}-y^{2}}$ for
(a)(c)(e)(g) and $d_{xy}$ for (b)(d)(f)(h). The pairing symmetry of SC is $%
d_{x^{2}-y^{2}}$ for (a)(b)(e)(f) and $d_{xy}$ for (c)(d)(g)(h). The rest of parameters are the same as those in \cref{Fig2}. }
\label{Fig3}
\end{figure*}
\end{widetext}
with $a_{he,s}$ ($a_{eh,s}$) being the coefficient
of Andreev reflection from incident electron (hole) to reflected hole
(electron) with spin-$s$. Here, we have made analytical continuation of
incident quasiparticle energy $E\rightarrow i\omega _{n}$ into Matsubara
frequencies $\omega _{n}=\pi k_{B}T(2n+1),(n=0,\pm 1,\pm 2....)$. Eq. \ref%
{FT} allows us to directly calculate the dc Josephson current in even more
complicated or long junctions, but we will focus on the short junction with $%
k_{F}L\ll \mu /\Delta $ in this work.

\section{Identification of dGSJ states via conductance}
\label{sec3}

First, we show the angle-resolved conductance $\sigma (\theta )$ and the
normalized conductance  $\sigma /\sigma _{0}$ of N/AM/$d$-SC junctions in \cref{Fig2} using Eq. \ref{cond1}. It is noted
that a clean contact between AM and $d$-SC enables Andreev reflection,
therefore we assume no barrier at the right interface. To observe the dGSJ
states, we require a finite barrier strength between N and AM (we set $Z=2$), which can confine the quasiparticle transport inside the AM. The length of
the AM layer is set to $k_{F}L=10$, which experimentally relates to the
short junction limit $k_{F}L\ll \mu /\Delta $. From Figs.~\ref{Fig2}(a) and \ref{Fig2}(c) we can
see that there are subgap resonance spikes when the altermagnetic order has $%
d_{x^{2}-y^{2}}$-wave symmetry, indicating the formation of dGSJ
states. As a result, the characteristic subgap resonance peaks show up in the normalized
differential conductance $\sigma /\sigma _{0}$, see Figs.~\ref{Fig2}(e) and \ref{Fig2}(g). It
is also interesting to note that in the short N/N/SC junction, dGSJ states
can hardly form due to the suppressed phase accumulations. However, as
explained in the introduction, the phase accumulation is greatly enhanced by
the $d_{x^{2}-y^{2}}$-AM order. Such induced dGSJ states have strong
dependence on $k_{y}$ and when the dispersion $E\left( k_{y}\right) $ is
almost flat across a certain range of $k_{y}$, the resonant peak in the
conductance $\sigma $ clearly appears. In comparison, a short $d_{xy}$-AM can not
produce the dGSJ states or the resonance spikes in the angle-resolved
conductance spectra as shown in Figs.~\ref{Fig2}(b) and \ref{Fig2}(d). Instead, we found the
well-known V-shape conductance and the zero-biased conductance peak 
for $d_{x^{2}-y^{2}}$-SC and $d_{xy}$-SC, respectively, as shown in  Figs.~\ref{Fig2}(f) and \ref{Fig2}(h).

\begin{figure*}\includegraphics[width=16cm]{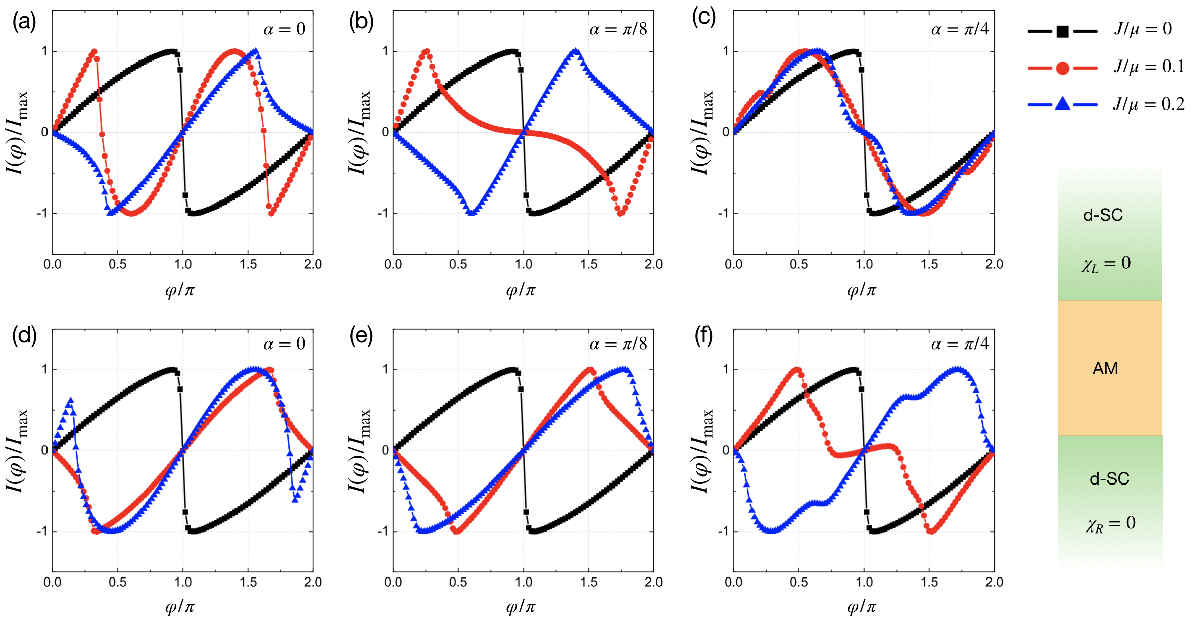}
\caption{ Current phase relation of $d_{x^2-y^2}$-SC/AM/$d_{x^2-y^2}$-SC Josephson junction. (a)(b)(c) $k_FL=10$ and (d)(e)(f) $k_FL=20$. We choose the temperature $k_BT=0.01084\Delta_0$. The current $I$
has been normalized to the critical current $I_{\max}=\max[I(\varphi)]$.  }
\label{nFig1}
\end{figure*}

To see the dependence of dGSJ states on the strength of the AM, we show the
density plot of the conductance spectra as a function of $eV$ and altermagnetic
strength $J$ in Figs.~\ref{Fig3}(a) to \ref{Fig3}(d). Figures~\ref{Fig3}(a) and \ref{Fig3}(c)
show that, with the increase of the strength of altermagnetism $J$, the
oscillatory conductance spikes emerge if AM has $d_{x^{2}-y^{2}}$-wave
order. Specifically, there are more subgap spikes for the $d_{x^{2}-y^{2}}$%
-SC junction than for the $d_{xy}$-SC junction for large value of $J$. This
indicates that $d_{x^{2}-y^{2}}$-AM order mainly suppresses the flat
zero-energy states in the $d_{xy}$-SC junctions. However, if the AM has $%
d_{xy}$-wave order, the characteristic of tunneling spectra is invariant as
compared to the junction without AM, as seen in 
Figs.~\ref{Fig3}(b) and \ref{Fig3}(d). We
further show the effect of the junction length on the emergence of dGSJ
states. Figs.~\ref{Fig3}(e) to \ref{Fig3}(h) show that the number of spikes is proportional to
$L$ since the phase accumulation increases with $L$. As expected, the dGSJ
states can be generated for the $d_{x^{2}-y^{2}}$-wave AM order in the short
junction (Figs.~\ref{Fig3}(e) and \ref{Fig3}(g)), but can only be found for the $d_{xy}$%
-wave AM order in the long junction limit [$k_F L>10^3$, see Figs.~\ref{Fig3}(f) to \ref{Fig3}(h)], because
of the negligible difference between electron and hole wave vectors. It is
worthwhile to point out that dGSJ states have already been shown in the
AM/AM/SC junction without unconventional pair potential \cite{Papaj23}  and thus found to be absent in semi-infinite AM/SC system \cite{Sun23} since the dGSJ states come from the
confinement effect \cite{Cayao21}.

\section{The Josephson effect}
\label{sec4}

In this section, we discuss the current phase relation (CPR) $I\left(
\varphi \right) $ in $d$-SC/AM/$d$-SC junctions, where $\varphi $ is the
phase difference between the left and and right pair potential. We set no
barrier between AM and SC, and leave the discussion of the barrier effect
elsewhere. Scattering coefficients are calculated numerically by imposing
boundary conditions for the scattering wave functions and the Josephson
current is obtained by Eq. \ref{FT}.

To analyze the CPR, we further decompose the
Josephson current into a series of different orders of Josephson coupling%
\begin{equation}
I\left( \varphi \right) =\sum\limits_{n}\left[ I_{n}\sin \left( n\varphi
\right) +J_{n}\cos \left( n\varphi \right) \right] ,
\end{equation}%
where $n$ is a positive integer. We consider different parameters such as
the crystal orientation $\chi _{L,R}$, $\alpha $ and find that the CPR is
expressed as $\sum\nolimits_{n}I_{n}\sin \left( n\varphi \right) $ and $J_{n}
$ is zero in our system. 
To demonstrate this, a relevant operator is the fourfold rotation
symmetry $C_{4}$ which corresponds to a rotation angle $\pi /2$ with respect
to $z$ axis and makes $k_{x}\rightarrow k_{y}$, $k_{y}\rightarrow -k_{x}$, $%
\hat{s}_{z}\rightarrow \hat{s}_{z}$, and $\varphi $ invariant. Another
relevant operator is the time-reversal symmetry $T$, which induces the
transformations $k_{x}\rightarrow -k_{x}$, $k_{y}\rightarrow -k_{y}$, $\hat{s%
}_{z}\rightarrow -\hat{s}_{z}$ and $\varphi \rightarrow -\varphi $. We
consider the combined symmetry $M_{0}=TC_{4}$ \cite{Bo2024} which is maintained in the system, 
\begin{equation}
M_{0}\hat{H}\left( \varphi \right) M_{0}^{-1}=\hat{H}\left( -\varphi \right) .
\end{equation}%
Consequently, the Josephson current satisfies the well-known characteristic $%
I\left( \varphi \right) =-I\left( -\varphi \right) $ with symmetry protected
zero net current at $\varphi =0$ or $\pi $, indicating that $J_n \cos \left( n\varphi \right)$ must vanish in the CPR of our Josephson junction.

\begin{figure*}[tbp]
\includegraphics[width=16cm]{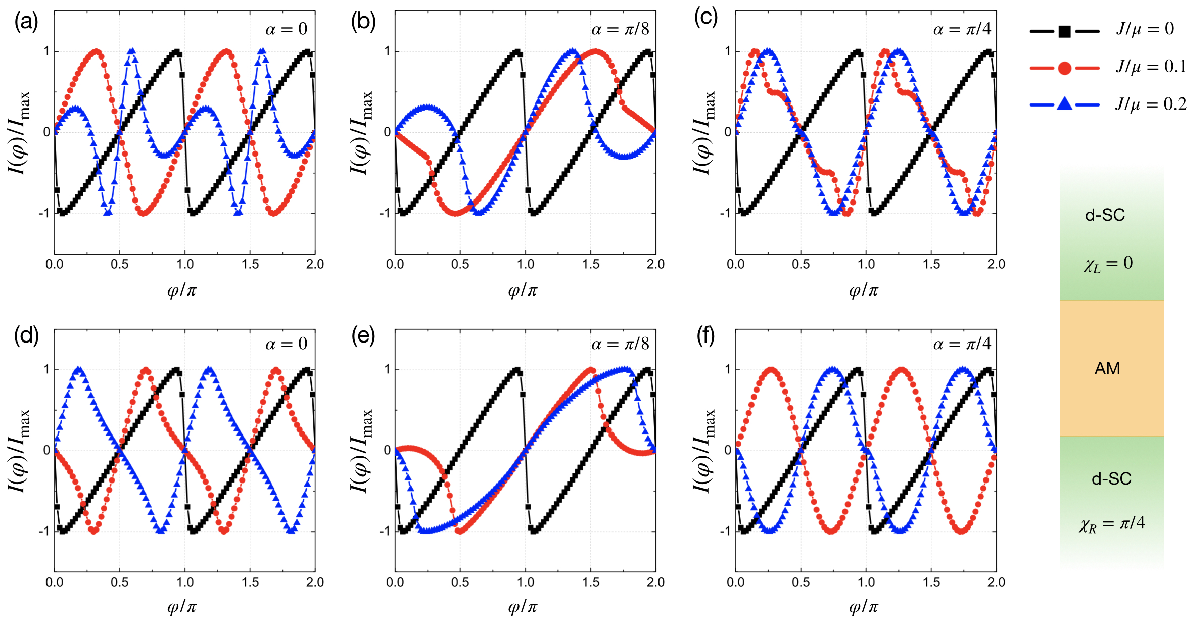}
\caption{ Current phase relation of $d_{x^2-y^2}$-SC/AM/$d_{xy}$-SC
Josephson junction. (a)(b)(c) $k_FL=10$ and (d)(e)(f) $k_FL=20$. We choose
the temperature $k_BT=0.01084\Delta_0$.}
\label{nFig2}
\end{figure*}


Figure \ref{nFig1} shows the Josephson current in $d_{x^{2}-y^{2}}$-SC/AM/$%
d_{x^{2}-y^{2}}$-SC junctions with various crystal orientations of the AM. It can
be seen that the altermagnetic strength $J$ can drive the $0-\pi $
transitions. Moreover, high-order Josephson coupling can be dominant in this
geometry, for example, the red curve in \cref{nFig1}(a). We also find
the anomalous skewness in the CPR, e.g., the blue curve in \cref{nFig1}(a)
and the red curve in \cref{nFig1}(d). Comparing to the upper panels (a)(b)(c) with lower panels (d)(e)(f), one can see that the $0-\pi $ transition can occur when the junction length varies, for example, the red curves in (b) and (e). The feature of Josephson current is also sensitive to the crystal orientation $\alpha $ of the AM, that is, $\alpha $ can also drive the $0-\pi $ transition when other parameters are kept the same. We find a similar behavior of the CPR in a $d_{xy}$-SC/AM/$d_{xy}$-SC junction.

We next show the CPR in asymmetric $d_{x^{2}-y^{2}}$-SC/AM/$d_{xy}$-SC
junctions in \cref{nFig2}. For $d_{x^{2}-y^{2}}$-AM, the second order
Josephson coupling is dominant but the sign of $I_{2}$ is highly tunable by
the strength of altermagnetism as shown in Figs.~\ref{nFig2}(a) and \ref{nFig2}(d). It is also
noted that the current at phase difference $\varphi =\pm \pi /2$ becomes
zero. Such behavior is the same with the case without magnetization. To
explain the nodal point at $\varphi =\pm \pi /2$, we consider the mirror
reflection with respect to $xz$-plane $M_{xz}$, which makes $%
k_{x}\rightarrow k_{x}$, $k_{y}\rightarrow -k_{y}$, $\hat{s}_{z}\rightarrow -%
\hat{s}_{z}$ and additional phase $\varphi \rightarrow \varphi -\pi $. We
use the magnetic mirror reflection symmetry $%
M_{1}=TM_{xz}$ \cite{Yamakage13,Boprb2015}, which leads to $k_{x}\rightarrow -k_{x}$, $k_{y}\rightarrow
k_{y}$, $\hat{s}_{z}\rightarrow \hat{s}_{z}$ and $\varphi \rightarrow
-\varphi +\pi $. We thus have%
\begin{equation}
M_{1}\hat{H}\left( \varphi \right) M_{1}^{-1}=\hat{H}\left( -\varphi +\pi \right) ,
\end{equation}%
for $d_{x^{2}-y^{2}}$-AM. As a result, $-I\left( -\varphi +\pi \right)
=I\left( \varphi \right) $ will be satisfied and we have $I\left( \varphi
=\pm \pi /2\right) =0$. It then excludes the presence of odd-order Josephson
coupling $I_{n}$ in the CPR. For $d_{xy}$-AM, we find a similar feature as
compared to $d_{x^{2}-y^{2}}$-AM. However, we need to use the complicated
combined operator $M_{2}=TM_{xz}C_{4}$ to explain the nodal point at $%
I\left( \varphi =\pm \pi /2\right) =0$, where $M_{2}$ makes $%
k_{x}\rightarrow k_{y}$, $k_{y}\rightarrow k_{x}$, $\hat{s}_{z}\rightarrow
\hat{s}_{z}$ and $\varphi \rightarrow -\varphi +\pi $. We arrive at
\begin{equation}
M_{2}\hat{H}\left( \varphi \right) M_{2}^{-1}=\hat{H}\left( -\varphi +\pi \right),
\end{equation}
for $d_{xy}$-AM and the system still has vanishing current $I\left( \varphi
=\pm \pi /2\right) =0$, as shown in Figs.~\ref{nFig2}(c) and \ref{nFig2}(f). For an AM that is
neither $d_{x^{2}-y^{2}}$-wave nor $d_{xy}$-wave, there is no operation that
maps $\hat{H}\left( \varphi \right) $ to $\hat{H}\left( -\varphi +\pi \right) $ and thus
the node at $\varphi =\pm \pi /2$ is no longer protected and found to be
lifted. Indeed, our numerical results are consistent with our symmetry
analysis since the first order Josephson coupling exists for $\alpha =\pi /8$
as shown in Figs.~\ref{nFig2}(b) and \ref{nFig2}(e). We conclude that the CPR, as well as the
tunneling conductance, sensitively depends on the orientation of the AM crystal.
This is in sharp contrast to the results in $d_{x^{2}-y^{2}}$-SC/ferromagnet/$%
d_{xy}$-SC junctions where the first-order sinusoidal component never appears in the CPR \cite%
{TanakaJPSJ2000}.

\section{Summary}
\label{sec5}

In summary, we have theoretically studied the differential conductance and Josephson effect in $d$-wave altermagnet/$d$-wave superconductor hybrids. 
We find that the subgap states, known as de Gennes-Saint-James states, can be enhanced by the $d_{x^{2}-y^{2}}$-altermagnet in the short junction but can only be formed in the long junction if the altermagnet has $d_{xy}$-wave order. We have shown that a robust zero-bias peak against the altermagnetic field can appear when both altermagnetic and superconducting order have $d_{xy}$-wave symmetry. We further reveal that the $0$-$\pi $ transition can occur in symmetric $d$-wave Josephson junctions by altermagnetism, which has been widely reported in conventional $s$-wave Josephson junctions. Notably, we find the orientation-dependent first order Josephson coupling in an asymmetric $d_{x^{2}-y^{2}}$-superconductor/altermagnet/$d_{xy}$-superconductor junction. This feature does not occur in Josephson junctions with $d$-wave superconductors and ferromagnets, unveiling a unique effect of altermagnetism. 
Since subgap states are known to promote the formation of odd-frequency spin-triplet Cooper pairs \cite{tanaka2012review,cayao2019odd,tanaka2024review}, our results suggest an intriguing possibility for enhancing emerging pair correlations in altermagnets \cite{Maeda2025,fukaya2024} and also a powerful way to control them via the Josephson effect. Our findings thus demonstrate the peculiar role of altermagnets on the Josephson effect, of practical significance for both controlling the Josephson current and designing new functional devices in superconducting spintronics.

\section{Acknowledgements}
Y. F. acknowledges financial support from the Sumitomo Foundation.
P. B. acknowledges support by the Spanish CM ``Talento Program'' project No.~2019-T1/IND-14088 and No.~2023-5A/IND-28927, the Agencia Estatal de Investigaci\'on project No.~PID2020-117992GA-I00 and No.~CNS2022-135950 and through the ``María de Maeztu'' Programme for Units of Excellence in R\&D (CEX2023-001316-M). 
J. C. acknowledges  financial support from the Swedish Research Council  (Vetenskapsr\aa det Grant No.~2021-04121) and the Carl Trygger’s Foundation (Grant No. 22: 2093).   
Y. T. acknowledges financial support from JSPS with Grants-in-Aid for Scientific research  (
KAKENHI Grants Nos.\ 23K17668, 24K00583, and 24K00556). 
B. L. acknowledge financial support from the National Natural Science Foundation of China (project 12474049).
	
\bibliography{altermagnet}

\end{document}